\documentclass{article}

\usepackage{graphicx}
\usepackage{epsfig} 
\newcounter{firstbib}

\usepackage{times}
\usepackage[superscript]{cite}

\usepackage[margin=0.9in]{geometry} 

\begin{document}


\title{Enhanced optical trapping via structured scattering}

\author{Michael~A~Taylor$^{1,2*}$, Muhammad Waleed$^{1}$, Alexander~B~Stilgoe$^{1}$, Halina Rubinsztein-Dunlop$^{1,3}$, \\
 and Warwick~P~Bowen$^{1,3}$ \vspace{3mm}\\ 
$^1$School of Mathematics and Physics, University of Queensland, St Lucia, Queensland 4072,\\
 Australia \vspace{1mm}\\
$^2$Research Institute of Molecular Pathology (IMP), Max F. Perutz Laboratories \& Research \\
Platform for Quantum Phenomena and 
Nanoscale Biological Systems (QuNaBioS), University of\\
 Vienna, 
Dr. Bohr Gasse 7-9, A-1030 Vienna, Austria \vspace{1mm}\\
$^3$Australian Centre for Engineered Quantum Systems, University of Queensland, St Lucia, \\
 Queensland 4072, Australia\vspace{3mm}}

\maketitle


{\bf 
Interferometry can completely redirect light, providing the potential for strong and controllable optical forces.
 However, small particles do not naturally act like interferometric beamsplitters, and the optical scattering from them is not generally thought to allow efficient interference.
Instead, optical trapping is typically achieved via deflection of the incident field. Here we show that a suitably structured incident field can achieve beamsplitter-like interactions with scattering particles. The resulting trap offers order-of-magnitude higher stiffness than the usual Gaussian trap in one axis, even when constrained to phase-only structuring.
 We demonstrate trapping  of 3.5 to 10.0~$\mu$m silica spheres, achieving stiffness up to 27.5$\pm$4.1 times higher than is possible using Gaussian traps, and two orders of magnitude higher measurement signal-to-noise ratio. These results are highly relevant to many applications, including cellular manipulation\cite{Thalhammer2011,Bowman2011}, fluid dynamics\cite{Franosch2011,Jannasch2011}, micro-robotics\cite{Palima2012}, and tests of fundamental physics\cite{Kheifets2014,Li2011}.
}

\vspace{3mm}

Optical forces are exerted when a particle changes the propagation direction of light. In optical tweezers this typically involves {\it deflection} of the light in proportion to the particle displacement, with minimal changes in its spatial structure (Fig.~\ref{Schematic}{\bf a}). Similarly, a beamsplitter which combines two fields will also experience an optical force (Fig.~\ref{Schematic}{\bf b}). In this case, the force arises because a relative phase shift between the incident fields redirects power between the outputs. Since the phase changes with beamsplitter displacement, interferometry can be used to stably trap.  Here, in contrast to deflection-based trapping, the interference redirects light between two discrete propagation directions. Displacements as small as $\lambda/(4 n_m)$ are capable of completely rerouting the light, where $\lambda$ is the wavelength and $n_m$ the medium refractive index. This far greater sensitivity can in principle allow trap stiffness that exceeds all previous experiments\cite{Jannasch2012} by over an order of magnitude. However, the beamsplitters which are traditionally used to interfere fields are poorly suited to optical micromanipulation experiments. Such experiments instead rely almost exclusively on scattering particles.

Recent works have shown that precise knowledge of scattering together with wavefront shaping allows particles to act as near-arbitrary optical elements, such as a mirror or lens~\cite{Katz2012}. In a similar way, we use non-trivial spatial structure in Mie scattering to convert  particles to effective beamsplitters which  separate a phase-structured input field into distinct output fringes (Fig.~\ref{Schematic}{\bf c}). 
Rather than deflecting the fringes, particle displacements act to redirect optical power between them. This enables a new form of beamsplitter-like optical trapping, which we term enhanced trapping via structured scattering (ENTRAPS).

To understand the forces achieved with ENTRAPS, it is useful to first review the force on an interferometer beamsplitter. When the incident fields have equal power $P/2$ and phase difference $\phi$, the power imbalance $\Delta P$ between the two output ports is $\Delta P = P \,{\rm cos} \phi$. This results in an optical force
\begin{equation}
F = \frac{P \, n_m}{c} \,{\rm cos} \phi \, {\rm sin} \theta, \label{F_beamsplitter}
\end{equation}
with $c$ the speed of light and $\theta$ the incident angle on the beamsplitter. Lateral movement of the beamsplitter by a distance $x$ causes a relative phase shift $\Delta\phi=4\pi n_m (x/ \lambda) \, {\rm sin} \theta$ between the input fields, and changes the force on the beamsplitter. If the initial phase is set to $\phi=\pi/2$ and the light is normally incident ($\theta=\pi/2$), this achieves a stable trap of the form $F=-\kappa x$, with trap stiffness $\kappa = \frac{4 \pi n_m^2 P}{c \lambda}$. As discussed in the supplementary information, without recourse to non-propagating fields or quantum states of light that exhibit non-classical photon correlations, this represents a fundamental upper limit on the achievable stiffness of any optical trap. 

%
%
ENTRAPS achieves strong trapping through similar principles, though instead using a scattering microparticle to interfere the fields. The particle can have any geometry, though here we restrict our discussion to homogeneous dielectric spheres for which the scattered light is described with Mie theory. It is well known that Mie scattering preferentially occurs into fringes at distinct scattering angles. However, the intensity of these fringes is usually much lower than the forward-scattering. Furthermore, neighbouring fringes are $\pi$~rad out-of-phase and destructively interfere if the incident illumination is homogeneous over a broad angular range, as is the case for a Gaussian trap. This further suppresses the fringe intensities (see Fig.~\ref{Schematic}{\bf a} (middle)).  As a consequence, Mie scattering fringes can usually be ignored in optical traps. However, we show here that they can be controllably populated  using structured illumination (Fig.~\ref{Schematic}{\bf c} (middle)). As such, the incident phase controls interference in Mie scattering fringes in a similar manner to the output of a beamsplitter. In ENTRAPS, the fringes are populated with fields that have near-orthogonal phase, such that a small phase shift can lead to constructive or destructive interference and redirect light between interference lobes on either side of the particle. This provides a beamsplitter-like trapping force (see Fig.~\ref{Schematic}{\bf c} (bottom)).

One might expect that full control of both the amplitude and phase is required to achieve ENTRAPS. However, the phase alone determines whether interference is constructive or destructive, while the amplitude profile determines the efficiency of the interference. As such, ENTRAPS can be achieved using phase-only control of the incident light. Henceforth in this paper, we focus on these phase-only implementations.
 While the restriction to phase-only control can be expected to limit the achievable stiffness, it makes the protocol easily accessible to widely used holographic optical tweezers  and, as we show here, still allows order-of-magnitude enhancements.

To identify appropriately structured incident optical fields for phase-only ENTRAPS and predict the performance of the protocol, we numerically optimised the phase profile of the incident field to maximize trap stiffness (see Methods and Supplement section S3). Particles with radius $R>\lambda$ were found to converge on profiles that achieve ENTRAPS, as evidenced by a particle-position-dependent redirection of power between distinct scattering fringes with fixed propagation directions (see polar plots in Fig.~\ref{SIFTscaling}{\bf a} and Supplementary section S3). We therefore postulate that, for particles that exhibit significant Mie scattering fringes, any trapping field that is near-optimized for trap stiffness will exhibit  ENTRAPS.

In general, Mie scattering profiles become increasingly complex as the particle size increases. This is apparent in the transmitted intensity patterns of ENTRAPS, which 
 feature an increasing number of bright fringes as the particle size increases (Fig.~\ref{SIFTscaling}{\bf a}). ENTRAPS relies on this complexity, and uses it to control the transmitted wavefront. Consequently, we find that the achievable enhancement in trap stiffness increases with the particle size, and exceeds an order of magnitude for sizes above 5~$\mu$m (Fig.~\ref{SIFTscaling}{\bf a}). A drop in stiffness enhancement is evident above 8~$\mu$m, which may indicate that the calculated holograms are not fully optimal. For particles between 7 and 10~$\mu$m in size, the predicted trap stiffness exceeds 3~mN~m$^{-1}$W$^{-1}$.   This
  is comparable to the current record demonstrated with anti-reflection coated high refractive index particles\cite{Jannasch2012}, even though the particles considered here are silica and thus low contrast.

 To test the predictions described above we implemented a holographic optical tweezer (see Fig.~\ref{Layout} and methods).
Silica microspheres  were trapped either with Gaussian input light or with the pre-calculated ENTRAPS holograms. 
 Thermal position fluctuations of the trapped beads were monitored using a position sensitive detector (PSD). Freely moving beads in water exhibit Brownian motion, which has a mechanical power spectrum that scales  with the inverse-square of frequency. The optical trap  suppresses low frequency motion, resulting in a spectrally flat region at low frequency\cite{Bowman2011}. The corner frequency quantifies the transition between these two spectral regions, and is directly proportional to the trap stiffness $\kappa$. The ENTRAPS enhancement factor could therefore be determined
 by comparing the corner frequencies obtained with ENTRAPS and Gaussian traps.

Experiments were performed with particles of 3.48~$\mu$m, 5.09~$\mu$m, 7.75~$\mu$m, and 10.0~$\mu$m diameter. These sizes are
relatively large for optical tweezers, with 10~$\mu$m near the upper limit for particles that are regularly trapped\cite{Palima2012,Mitchem2008}. 
To illustrate these measurements, Fig.~\ref{SIFTscaling} shows mechanical power spectra measured with 7.75~$\mu$m particles. The corner frequency measured with ENTRAPS is 73.2$\pm2.8$~Hz, compared to 2.66$\pm0.38$~Hz for the Gaussian trap (Fig.~\ref{SIFTscaling}{\bf b} and {\bf c}). This constitutes an increase in trap stiffness by a factor of 27.5$\pm4.1$, consistent with the predicted enhancement of 26.1. The enhancement was close to theory for all particle sizes, as shown in Fig.~\ref{SIFTscaling}{\bf a}, with mechanical spectra given in Supplementary section S4.

In addition to improving the trap stiffness, ENTRAPS also provides a dramatic improvement in the measurement signal-to-noise ratio (SNR). Laser tracking in optical tweezers is based on measurements of the centroid position of the transmitted light. This centroid is proportional to the applied optical force. As such, the observed signal at a given displacement is enhanced as the trap stiffness increases.  When using 7.75~$\mu$m particles, ENTRAPS improved the SNR by a factor of 249, which allowed the thermal motion of the particle to be measured with an order of magnitude higher bandwidth (see Fig.~\ref{SIFTscaling}{\bf b} and {\bf c}). Substantial SNR enhancements were observed for all particle sizes, and exceeded two orders of magnitude for both 5.09 and 10.0~$\mu$m beads.

ENTRAPS complements a range of novel approaches to optical micromanipulation that have only recently been demonstrated. It has been shown that the direction of optical forces need not align with the propagation of light, including a lateral force applied with Airy beams\cite{Baumgartl2008} and pulling force applied with a Bessel beam\cite{Chen2011} and an interference-based tractor beam\cite{Brzobohaty2013}. Recently, negative torque which opposes the optical angular momentum has also been demonstrated\cite{Hakobyan2014}. 

While it has been known for over two decades that structured light fields can be used to improve optical traps\cite{Ashkin1992}, no previous proposal or experiment has made use of the spatial structure of scattering to improve trap stiffness. As such, ENTRAPS represents a fundamentally different approach. Furthermore, compared to the order-of-magnitude enhancements possible with ENTRAPS, previous experiments have only allowed relatively modest enhancements. Higher order Laguerre-Gaussian modes have been shown to increase the axial restoring force on 5~$\mu$m particles by a factor of 1.60\cite{ONeil2001};
while radially and axially polarized modes have been predicted to provide even better trapping strengths\cite{Nieminen2008}, with experimental realisations enhancing the axial and lateral trapping forces by factors of 1.30 and 1.16, respectively\cite{Kozawa2010}. 
 
To date the most powerful method to improve trapping forces has been to engineer the particle rather than structure the light. Most notably, anti-reflection coated high refractive index titania particles allow
 transverse trapping forces double that achieved with polystyrene microspheres\cite{Jannasch2012}, which remain the strongest trapping forces and trap stiffness ever reported. The axial trapping was simultaneously improved, though with a smaller (1.3-fold) enhancement.
Alternatively, back-scatter can be minimized for a homogeneous sphere by careful choice of diameter, which also improves trap stiffness\cite{Stilgoe2009}. Another recent demonstration achieved strong static optical forces, though not optical trapping, on a bent waveguide structure that
redirects the light with near-perfect efficiency\cite{Palima2013}.

Even while achieving substantial enhancements in trap stiffness and measurement signal-to-noise, we note that our currently implementation of ENTRAPS has significant limitations. Firstly, compared to the three-axis enhancement achieved, for example, in Ref.~\cite{Jannasch2012}; its achieves enhancement only along a single axis. 
It is possible that this could be overcome in future using three-axis-optimised phase profiles.
 Secondly, it can be seen from Fig.~\ref{Schematic}{\bf a}~and~{\bf c} (bottom) 
 that ENTRAPS has a reduced trapping region compared to a Gaussian trap. In the case of a 10.0~$\mu$m particle, for instance, our theory and experimental force calibration (see Fig.~4 in the Supplementary information) show a reduction  from 10~$\mu$m to approximately 500~nm. Indeed, a general trade-off exists between trap stiffness and trapping region. 
 The  scattering force on a trapped particle initially increases linearly with displacement in proportion to the trap stiffness.
 However, since photon scattering events cannot exert transverse momentum kicks larger than $\hbar k$ this cannot continue indefinitely, with the region of linear response reducing as the trap stiffness increases. In our experiments the momentum kick per photon reaches a maximum of approximately $\hbar k/7$. It is therefore concievable that the trapping region could be enlarged by a factor of seven without compromising stiffness. 
Finally, due to the presence of intricate phase structure ENTRAPS favours particles with volumes large enough to accommodate interference. ENTRAPS is therefore best achieved using particles larger than an optical wavelength.

Optical trapping has many important applications for particles larger than the wavelength. However, these applications are much less well explored than the small particle regime, as Gaussian traps lose trap stiffness with increasing diameter\cite{Bowman2013} (see Fig.~\ref{SIFTscaling}{\bf a} (inset)). For instance, cells are often manipulated optically, though in many cases they are too large to efficiently use a single-beam optical trap and require counter-propagating fields\cite{Thalhammer2011,Bowman2011}. Likewise, aerosols up to 10~$\mu$m are regularly trapped and studied in optical tweezers to gain insights into pollution, vapour based drug delivery, and atmospheric physics~\cite{Mitchem2008}. Micro-robotics also often relies on optical control of large spherical particles; either as handles that are attached to larger compound structures, or with the microsphere itself acting as the micro-robot\cite{Palima2012}.
 Microspheres are also levitated in vacuum,
where high mass can be important in studies of gravitational forces and particle physics\cite{Li2011,Moore2014,GeraciPRL}. The improved stiffness and measurement precision of ENTRAPS could greatly improve such applications.

Studies of hydrodynamics could also greatly benefit from ENTRAPS, since large particles couple more strongly to fluid flow\cite{Kheifets2014,Jannasch2011}. For instance, particle motion is expected to decouple from its surrounding fluid envelope only at very short time-scales, beyond the reach of state-of-the-art technology for a 1~$\mu$m sphere\cite{Huang2011} but accessible with ultraprecise tracking of 10~$\mu$m particles. Hydrodynamic resonances are predicted to allow coherent energy exchange 
between the particle and fluid flow.
 Such resonances have recently been observed\cite{Franosch2011,Jannasch2011}, though with trap stiffness a factor of five beneath 
 the strong resonance condition. 
 ENTRAPS may allow this condition to be met, which would provide a test of hydrodynamic theory in a previously uncharted regime. The highly localized fluid flow generated near the particle could also provide a new avenue for non-contact and low-damage manipulation of cells, similar to experiments utilizing fluid flow near rotating particles\cite{Ye2014}. It may also allow mechanical sensing of the fluid properties within the  localized flow region, thus allowing nanoscale microrheology\cite{Franosch2011}.  We envision that ENTRAPS could greatly improve these applications and others like them, and thus play an important part in the future of optical manipulation.


\section*{Acknowledgements}
 This work was supported by the Australian Research Council Discovery Project Contract No. DP140100734, and by the Air Force Office of Scientific Research Grant No. FA2386-14-1-4046. W.P.B. acknowledges support through the Australian Research Council Future Fellowship scheme FF140100650.

\section*{Competing Interests}
The authors declare that they have no
competing financial interests.

\section*{Correspondence} Correspondence and requests for materials
should be addressed to M.A.T.~(email: michael.taylor@imp.ac.at).

\section*{Author contributions} M.A.T. and W.P.B. conceived of and led the project. M.A.T. developed the theoretical concepts, and performed the calculations and analysis. A.B.S. and H.R.D. developed the experimental apparatus. M.W. and  M.A.T. performed the experiments, with assistance from A.B.S.. M.A.T. and W.P.B. wrote the paper with input from all coauthors.


\setcounter{figure}{0}

\section*{Figure captions}

\begin{figure}[h!]
 \begin{center}
   \includegraphics[width=8.5cm]{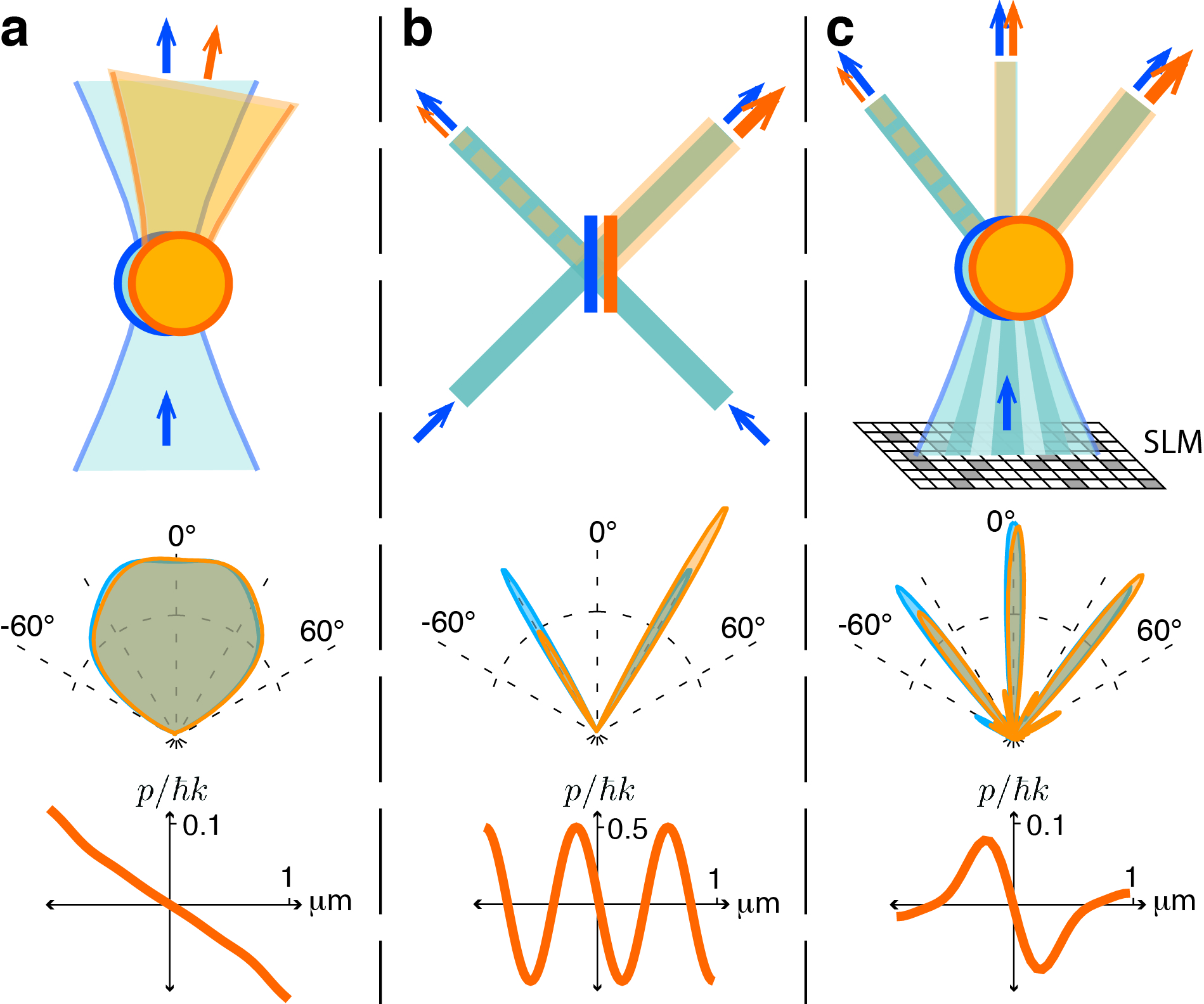}
   \caption{
 {\bf Trapping via Mie interference.} 
 Optical trapping 
 via deflection 
  is shown in {\bf a} both schematically (top) and with a calculated polar plot of the scattered intensity (middle). Deflections change the transverse momentum $p$ of scattered photons (bottom) where 
  $k=2\pi/\lambda$, inducing a transverse position dependent optical force. 
 {\bf b} Interference  
 at a beamsplitter
 can achieve a far stiffer trap, though with a smaller trapping range (bottom). Here, the polar and optical momentum plots are calculated for near-collimated Gaussian inputs at 60$^\circ$ 
 incidence.
  {\bf c} ENTRAPS applies a similar force on a microparticle.
   A structured trapping field is split into distinct scattering fringes,
    with particle displacement rerouting power between fringes. 
 {\bf a} and {\bf c} are calculated for a 3~$\mu$m silica sphere in water, with polar plots (middle) shown both for a particle that is centred (blue) and displaced by 150~nm (orange). In  {\bf b} the displacement is 50~nm.
}
 \label{Schematic}  
 \end{center}
\end{figure}

\begin{figure}[h!]
 \begin{center}
   \includegraphics[width=8.5cm]{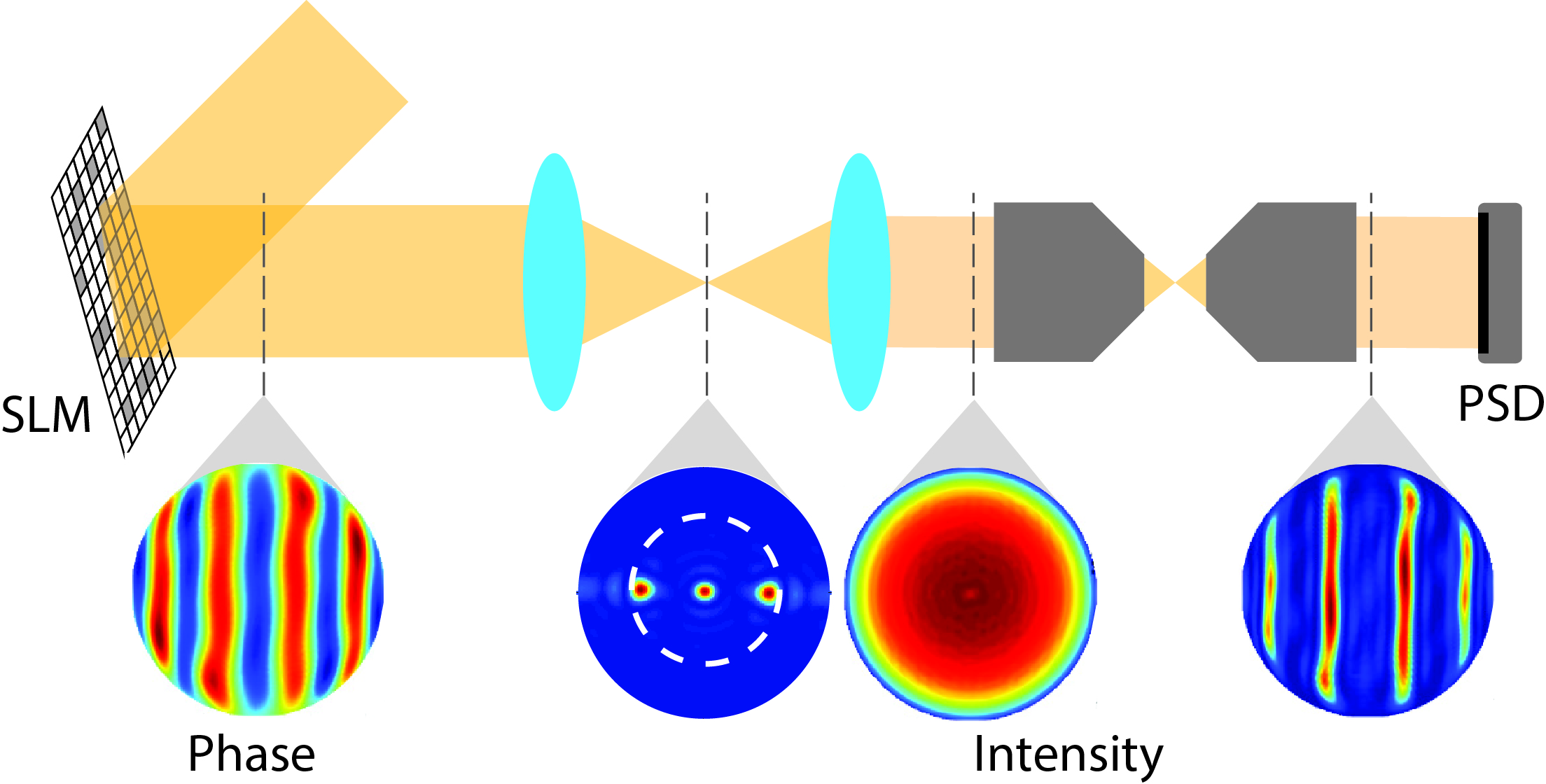}
   \caption{
 {\bf Layout of experiment.} ENTRAPS is implemented here using holographic optical tweezers, with a spatial light modulator (SLM) used to engineer the phase of the trapping field. The insets show the applied phase and resulting intensity profiles calculated for 3.48~$\mu$m diameter particles. In the image plane this field shows three distinct peaks and associated curved fringes, though it maintains a Gaussian intensity distribution at the back-focal plane. Interaction with the particle separates the light into discrete vertical fringes which each have a position-dependent power. The phase, back-focal plane, and output profiles are scaled to the back aperture width of an NA1.25 objective; while the dashed circle indicates the effective particle diameter in the image plane.
}
 \label{Layout}  
 \end{center}
\end{figure}

 \begin{figure*}[h!]
 \begin{center}
   \includegraphics[width=12cm]{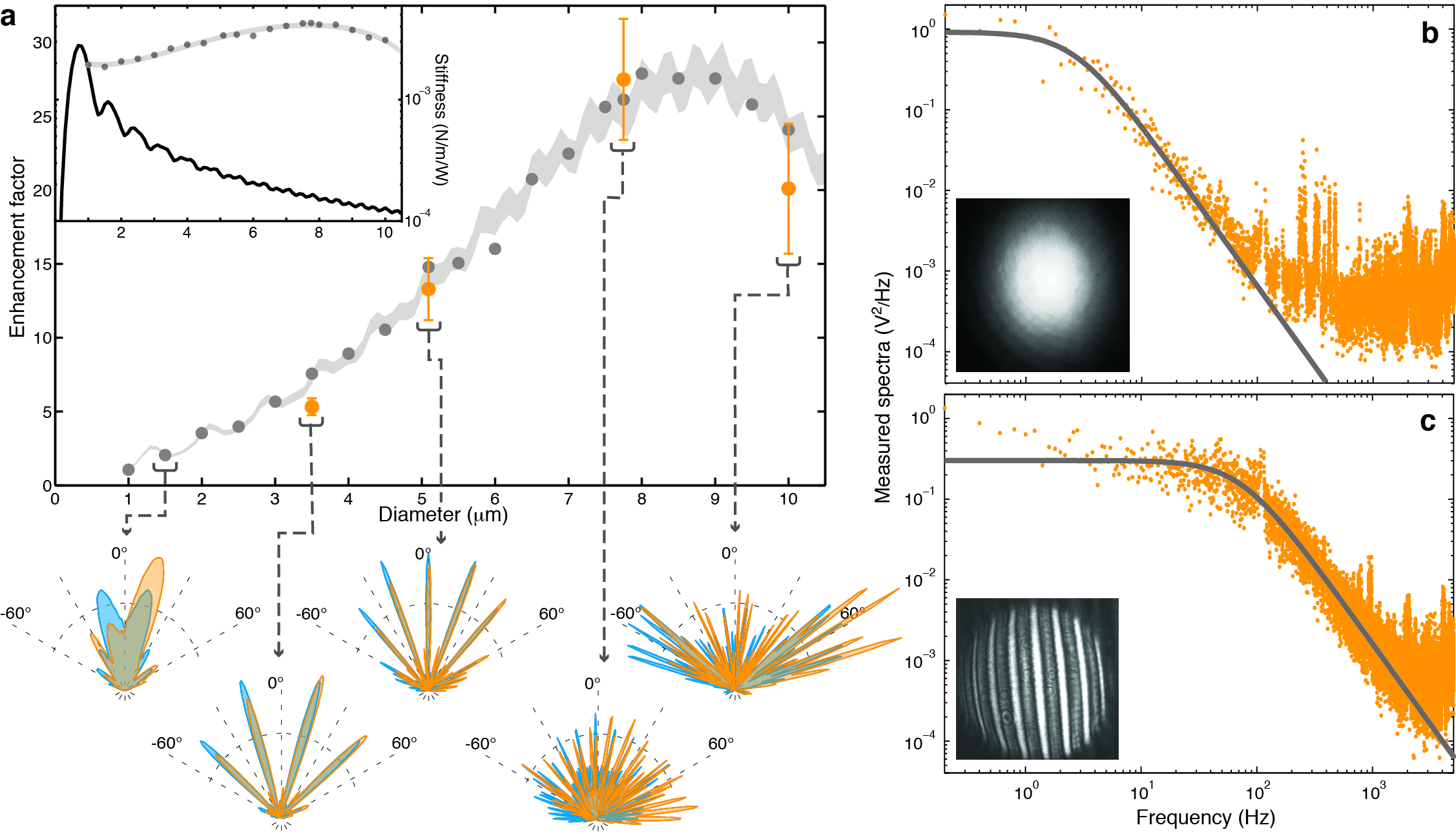}
   \caption{
 {\bf Trap stiffness of ENTRAPS.} {\bf a}: Stiffness enhancement factor as a function of particle size. Grey points: simulations, orange points: experiment. Error bars were determined statistically 
  (see supplementary information). Inset: Simulated
  stiffness of Gaussian (black curve) and ENTRAPS (grey points) traps.
   Grey band: three-parameter polynomial fit to the ENTRAPS simulation, with
 width
  equal to 5\% of the stiffness defining a rough uncertainty window for the simulation. The grey band in the main figure is derived from this band.
Polar plots:
 angular scattering profile of ENTRAPS on a linear scale for particle sizes of $1.5$, $3.48$, $5.09$, $7.75$, and $10.0~\mu$m from left to right, respectively. Blue (orange) shading labels profiles for particles that are centred (offset by $150$~nm). 
 $\bf b$ and $\bf c$: 
Mechanical power spectra for a $7.75~\mu$m particle 
in Gaussian and ENTRAPS traps, respectively. Orange points: measured data,
grey curve: Lorentzian fit, insets: observed intensity distribution after the trap. 
 }
 \label{SIFTscaling}  
 \end{center}
\end{figure*} 

\newpage

\section*{Methods}

\subsection*{Optimisation algorithm} 

ENTRAPS phase profiles were determined using a ``method of steepest descent" algorithm, with trapping forces calculated using the Optical Tweezers Toolbox\cite{Nieminen2007}; as described in more detail in supplementary section S2. This algorithm deterministically converges on a local optima, though this may not correspond to the global optima.  More efficient optimization protocols are available. Most particularly, using the Optical Eigenmode Method it may prove possible to efficiently identify globally optimal profiles\cite{Mazilu2011,Mazilu2012}.
However, implementation of these profiles, in general, would
require arbitrary control over both phase and amplitude.
 It would be interesting in future to compare the global optima to the phase-only solutions used here, and thereby establish the additional benefit provided by amplitude control. We note that there is a very large parameter space to explore, and it is unlikely that ENTRAPS is the only new characteristic trapping behaviour that can be achieved with structured fields.

\subsection*{Effect of trapping beam polarisation} 

A surprising result is that the optimized phase holograms show some deviations from perfect mirror-symmetry, and are instead symmetric under a 180$^\circ$ rotation. This is because the optimization was performed for circular polarization, which breaks the mirror symmetry of the trapping light. The slight deviations from mirror symmetry are important to the performance of the trap. The hologram shown in Fig.~\ref{Layout} provides a 15\% higher trap stiffness when right-circular polarization, for which it was calculated, than left-circular polarized light. By comparison, left-circular polarized light performs best with the mirror image of the hologram shown; and both horizontal and vertical polarizations trap most efficiently with mirror-symmetric holograms.

\subsection*{Experimental implementation} 

The holographic optical tweezer used a reflective SLM (Holoeye HEO-1080P), that was  imaged onto the back-focal plane of the trapping objective with relay lenses to allow arbitrary phase control of the trapping light. The input optical power was kept constant between experiments, with 250~mW input at the back-focal plane of the trapping objective. The light was focused on a silica microsphere 
  using an oil-immersion objective (Zeiss EC-Plan Neofluar, NA1.3). After trapping the microsphere, the light was collected with a matching condenser objective (Zeiss EC-Plan Neofluar, NA1.3).
%
The calculated hologram was applied on the SLM along with a blazed grating to eliminate zero-order reflections, and an aberration compensation pattern which was determined in advance. All diffraction modes other than first order mode were eliminated with an aperture, while the first order mode propagates through the trap. The PSD used to determine the change in optical momentum was custom-made in-house.
A PSD is well suited for tracking particles in optical tweezers, and has the additional benefit of accurately measuring the applied optical force\cite{Farre2012}. The optical trap was also monitored with a CCD using LED illumination to visually identify particles.

\subsection*{Technical considerations} 

There are a number of complicating factors which might be expected to limit the trap stiffness of ENTRAPS. SLMs have phase flicker at their video refresh rate. This flicker has previously been characterized to have a 1~radian peak-to-peak amplitude when applying a 2$\pi$ phase shift to our SLM\cite{Tay2009}.
 Any misalignment or aberrations will also cause the trapping field to deviate from the ideal field described in the calculations. In the experiments reported here, aberrations were compensated with the SLM to visually optimize the point spread function of the Gaussian field, though it is likely that some aberrations remained. Despite these technical limitations, our demonstrated enhancement in trap stiffness was within experimental error of the prediction. This shows that ENTRAPS is remarkably robust and can be applied with relatively basic trapping equipment.

The phase flicker on the SLM
results in an intensity ripple on the trapping light, and a consequent ripple on the detected signal. This ripple is evident as a series of spectral peaks at integer multiples of 50~Hz.  The data displayed in Fig.~\ref{SIFTscaling}~{\bf b}~and~{\bf c} has been filtered to remove this ripple, and also some additional spectral peaks which correspond to mechanical resonances of the optical tweezers apparatus.

\end{document}